\documentclass[12pt]{amsart}
\usepackage{geometry}            
\usepackage{graphicx}
\usepackage{amssymb}
\usepackage{mathtools}
\usepackage{epstopdf}
\usepackage{color}
\usepackage{mathpazo}
\fontfamily{Palatino}
\newcommand{\e}{\mathrm{e}}

\newcommand{\be}{\begin{equation}}
\newcommand{\ee}{\end{equation}}
\newcommand{\ba}{\begin{eqnarray}}
\newcommand{\ea}{\end{eqnarray}}

\newcommand{\diff}{{\rm d}}

\newcommand{\QQ}{\mathbb{Q}}

\newcommand{\lp}{\left(}
\newcommand{\rp}{\right)}
\newcommand{\lb}{\left[}
\newcommand{\rb}{\right]}

\begin{document}

\begin{center}
\tiny{Essay written for the Gravity Research Foundation 2019
Awards for Essays on Gravitation}

\vspace{1.5cm}
\huge{\bf The canonical frame of purified gravity}
\vspace{1cm}

\large{\bf  Jose Beltr\'an Jim\'enez$^*$, Lavinia Heisenberg$^\dagger$ and Tomi S. Koivisto$^\ddag$}  \\
\vspace{1cm}
\end{center}
{\tiny{
$^*$Departamento de F\'isica Fundamental, Universidad de Salamanca, E-37008 Salamanca, Spain.\\
$^\dagger$ Institute for Theoretical Studies, ETH Zurich, Clausiusstrasse 47, 8092 Zurich, Switzerland.\\
$^\ddag$Nordita, KTH Royal Institute of Technology and Stockholm University, Roslagstullsbacken 23, 10691 Stockholm, Sweden; \\
Laboratory of Theoretical Physics, Institute of Physics, University of Tartu, W. Ostwaldi 1, 50411 Tartu, Estonia.}}
\vspace{1cm}
\begin{center}
{\large{\bf Abstract}}
\end{center}
\small{In the recently introduced gauge theory of translations, dubbed Coincident General Relativity, gravity is described with neither torsion nor curvature in the spacetime affine geometry.  
The action of the theory enjoys an enhanced symmetry and avoids the second derivatives that appear in the conventional Einstein-Hilbert action.
While it implies the equivalent classical dynamics, the improved action principle can make a difference in considerations of energetics, thermodynamics, and quantum theory. This essay reports on possible progress in those three aspects of gravity theory. In the so-called purified gravity, 1) energy-momentum is described locally by a conserved, symmetric tensor, 2) the Euclidean path integral is convergent without the 
addition of boundary or regulating terms and 3) it is possible to identify a canonical frame for quantisation.}

\date{\today}                                           


\newpage
\normalsize
The action principle for the General theory of Relativity (GR) that Einstein had written down in 1916 \cite{einstein16} is
\be \label{le}
S_E = \int \diff^4 x \sqrt{-g}L_E\,, \quad L_E=\frac{1}{16\pi G}g^{\mu\nu}\Big(\left\{^{\phantom{i} \alpha}_{\beta\mu}\right\} \left\{^{\phantom{i} \beta}_{\nu\alpha}\right\} -\left\{^{\phantom{i} \alpha}_{\beta\alpha}\right\}\left\{^{\phantom{i} \beta}_{\mu\nu}\right\} \Big)\,,
\ee
where $G$ is the Newton's constant, $\left\{^{\phantom{i} \alpha}_{\beta\mu}\right\}$ are the Christoffel symbols of the metric $g_{\mu\nu}$ and $g$ is its determinant. Though $S_E$ does not suffer from second derivatives of the metric, the action principle due to Hilbert, which is a volume integral over the curvature $\mathcal{R}$ of the Christoffel symbols, became the more generally accepted standard from early on. 

Einstein proposed \cite{einstein16} that the energy and momentum of the gravitational field is described by
\be \label{te}
t_E^\mu{}_\nu = \frac{\partial L_E}{\partial g_{\alpha\beta,\mu}}g_{\alpha\beta,\nu} - \delta^\mu_\nu L_E\,,
\ee
which is indeed obtained from the canonical Noether current corresponding to the invariance of $S_E$ under translations. This proposal was immediately critised \cite{cattani}, on the grounds
that $t_E^\mu{}_\nu$ is not a tensor but a pseudotensor. Thus, in the absence of
gravity it can be non-vanishing and in the presence of gravity it can be vanishing, depending on the coordinate system. Nevertheless, Einstein vigorously defended  \cite{cattani} the use of the pseudotensor (\ref{te}),
which was also adopted, amongst many others, by Dirac in his textbook \cite{dirac}. Though a myriad of non-canonical pseudotensors and quasilocal definitions of the gravitational energy-momentum have been introduced by many other authors (for brief reviews see e.g. \cite{Aguirregabiria:1995qz, nester, Xulu:2002ix}), it is fair to say that none of them provides a fully compelling alternative to the original proposal \cite{einstein16}. In this essay we shall revisit it from the covariant perspective of {\it purified gravity} \cite{BeltranJimenez:2017tkd,Koivisto:2018aip,BeltranJimenez:2018vdo}.


In the framework of purified gravity, gravitation is understood as an inertial (i.e. pure gauge) ``force'' rather than as spacetime geometry, and GR is reformulated as the gauge theory
of translations that was dubbed the Coincident GR (CGR) \cite{BeltranJimenez:2017tkd}. Though Einstein considered GR in terms of both curvature and torsion \cite{Goenner:2004se}, in his view the great achievement of the theory never was the geometrisation of gravitation {\it per se}, but its unification with inertia \cite{lehmkuhl}.  
Technically\footnote{Since the coframe $\e^a=\theta^a + Dx^a$ is the translation gauge potential $\theta^a$ only up to the covariant derivative $Dx^a$ of the Cartan radius vector $x^a$ \cite{Hehl:1994ue}, the torsion $T^a \equiv D\e^a = D\theta^a + R^a{}_bx^b$ is the translation field strength if teleparallelism, $R^a{}_b=0$, is assumed. On the other hand, in standard GR it holds instead that $T^a=0$, and the translation gauge field strength $D\theta^a=-R^a{}_bx^b$ is in turn directly proportional to the curvature.}, GR equally well as its teleparallel equivalent \cite{Aldrovandi:2013wha} can be viewed as a gauge theory of translations.
 However, in purified gravity the defining property of the covariant derivative $\nabla_\mu$ is its commutativity, $[\nabla_\mu,\nabla_\nu]=0$, which distinguishes purified gravity as the canonical framework for a gauge theory of the Abelian group 
 of translations. The affine connections corresponding to translations\footnote{In purified gravity translations are realised passively, as general coordinate transformations. Note that in contrast to the usual approach to metric-affine gauge theories \cite{Hehl:1994ue}, we do not gauge translations in addition to the general linear group but from within that group. To our knowledge, an equivalent (pure) gauge theory was first
 presented in \cite{Wallner:1981jf}, see \cite{Koivisto:2018aip,BeltranJimenez:2018vdo} for more references.}
  are precisely those with neither curvature nor torsion \cite{BeltranJimenez:2017tkd}. This implies that the field strength of translations vanishes, which is the gauge theoretical rationale underlying the equivalence principle \cite{Koivisto:2018aip}.

In CGR, the metric $g_{\mu\nu}$ represents the potential of the inertial ``force'' field $Q_{\alpha}{}^{\mu\nu} \equiv -\nabla_\alpha g^{\mu\nu}$. 
The action of the theory is singled out as the unique quadratic form that is invariant with respect to infinitesimal translations of the connection \cite{BeltranJimenez:2017tkd}. Including 
a matter Lagrangian $L_M$, this action can  be written as
\be \label{lq}
S = \int \diff^4 x \sqrt{-g}\lp\frac{1}{16\pi G}\QQ + L_M\rp\,, \quad \QQ \equiv -Q_{\alpha}{}^{\mu\nu}P^\alpha{}_{\mu\nu}\,, 
\ee  
where $\QQ$ is the quadratic invariant that is given by the constitutive tensor $P^\alpha{}_{\mu\nu}$ as
\be
P^\alpha{}_{\mu\nu} = -\frac{1}{4}Q^\alpha{}_{\mu\nu} + \frac{1}{2}Q_{(\mu\nu)}{}^\alpha + \frac{1}{4}\lp Q^\alpha - \tilde{Q}^\alpha\rp g_{\mu\nu} - \frac{1}{4}\delta^\alpha_{(\mu}Q_{\nu)}\,,
\ee
where we have defined the Weyl covector $Q_\alpha \equiv Q_{\alpha\mu}{}^{\mu}$ and the projective trace $\tilde{Q}^\alpha \equiv Q_{\mu}{}^{\alpha\mu}$.   Because the connection is integrable, it is always possible to trivialise it by the gauge choice called the {\it coincident gauge} \cite{BeltranJimenez:2017tkd}, $\mathring{\nabla}_\alpha=\partial_\alpha$ (we denote quantities in this gauge by placing a ring over the symbols). One can verify that the
theory described by $S$ in (\ref{lq}) is dynamically equivalent to GR by checking from (\ref{le}) that $\mathring{\QQ}=L_E$. 

The field equations obtained by the variation of the action $S$ with respect to the metric can be written as 
\be \label{meom}
\tau^\mu{}_\nu - t^\mu{}_\nu = T^\mu{}_\nu\,,
\ee
where we have defined the metric and the matter energy-momentum tensors as
\be \label{canonical}
t^\mu{}_\nu \equiv \frac{1}{8\pi G}\lp P^{\mu\alpha\beta}Q_{\nu\alpha\beta} + \frac{1}{2}\delta^\mu_\nu \QQ\rp\,, \quad T_{\mu\nu} \equiv \frac{-2}{\sqrt{-g}}\frac{\delta \lp \sqrt{-g} L_M\rp}{\delta g^{\mu\nu}}\,,
\ee 
respectively. In addition, there appears what we could call the inertial energy-momentum tensor \cite{BeltranJimenez:2018vdo}, given by 
\be \label{inertial}
8\pi G\tau^\mu{}_\nu \equiv -\frac{2}{\sqrt{-g}}\nabla_\alpha\left(\sqrt{-g} P^{\alpha\mu}{}_{\nu}\right)\,. 
\ee
With some tensor algebra, we can see that $\sqrt{-\mathring{g}}\mathring{\tau}^\mu{}_\nu$ is an expression for what is
known as the Einstein energy-momentum complex \cite{Aguirregabiria:1995qz,nester,Xulu:2002ix}, and by comparing (\ref{te}) and (\ref{canonical}) we can confirm that
$\mathring{t}^\mu{}_\nu = t_E^\mu{}_\nu$. Thus, CGR naturally offers the fully covariant improvement of the canonical split of the gravitational energy budget \cite{einstein16}.  

It is crucial to also take into account the equation of motion for the connection, which is now independent of the metric. 
Assuming for simplicity that the independent connection does not enter into $L_M$, the variation of $S$ in (\ref{lq}) yields \cite{BeltranJimenez:2018vdo}
\be \label{ceom}
\nabla_\mu \lp \sqrt{-g} \tau^\mu{}_\nu\rp = 0\,.
\ee
This precisely ensures the conservation of the inertial energy-momentum in (\ref{inertial}).
 The result (\ref{ceom}) can also be obtained as a geometrical Bianchi identity \cite{BeltranJimenez:2018vdo}. It holds in any frame and in particular, due to (\ref{meom}), implies that $\partial_\mu[\sqrt{-\mathring{g}}(\mathring{t}^\mu{}_\nu + \mathring{T}^\mu{}_\nu)]=0$.
\newline
\newline

We are now in a position suggest a covariant criterion for {\it the canonical frame}. The conjecture is that a canonical frame is defined by $t^\mu{}_\nu=0$, i.e. the vanishing of the 
energy-momentum associated with the spacetime metric. In such a frame the tensor $\tau^\mu{}_\nu$ describes the local energy-momentum of spacetime and matter. 
Some remarks may be in order.
\begin{itemize}
\item Hoping not to have lost the reader with too many new terms we briefly recall them. Purified gravity: pure gauge theory of translation. Coincident GR: The equivalent of GR in purified gravity. Coincident gauge: the unitary gauge wherein $\mathring{\nabla}_\alpha=\partial_\alpha$. Canonical frame: a frame wherein $t^\mu{}_\nu=0$. 
\item It is worth reiterating that the condition $t^\mu{}_\nu=0$ is covariant i.e. totally independent of the coordinate system. (Of course, if one insists on working in the coincident gauge, the gauge redundancy is eliminated and $\mathring{t}^\mu{}_\nu$ becomes coordinate-dependent.)
\item It is evident that a canonical frame always exists (and that one can always at least locally further stipulate that $\mathring{t}^\mu{}_\nu=0$). Thus, in CGR we can determine the gravitational 
energy-momentum at any given point of spacetime\footnote{This may not still quite amount to the statement \cite{bondi} that  ``In relativity a non-localizable form of energy is inadmissible because any form of energy contributes to gravitation and so its location can in principle be found.'' In Schwarzschild spacetime, to be discussed shortly, we have everywhere $\tau^\mu{}_\nu=0$ but still find nonzero energy charge. Perhaps it should understood that the energy resides at the singularity.}. 
\item Since in the canonical frame the field equations of CGR reduce to $\tau^\mu{}_\nu=8\pi G T^\mu{}_\nu$, our conjecture by construction incorporates the Cooperstock hypothesis \cite{cooperstock}, namely that energy only exists in regions where the matter energy-momentum tensor is non-vanishing. 
\item Because of the latter identity, the inertial energy-momentum tensor $\tau^{\mu\nu}=\tau^{(\mu\nu)}$ is symmetric in the canonical frame. Thus it avoids another problem of the original proposal \cite{einstein16} and facilitates the definition of the gravitational angular momentum. 
\item The concept that the metric is an auxiliary field that should have vanishing energy-momentum tensor also more and less strongly resonates with many ideas about quantum \cite{DeWitt:1967yk}
and emergent \cite{Sindoni:2011ej} gravity, from Sakharov's induced gravity to topological and holographic considerations in string theory. We believe that CGR could provide a fresh impetus to some of these ideas, but will have to enter into that discussion elsewhere.  
\end{itemize}

At this point of the essay it is pertinent to demonstrate that the conjecture works in practice. For this purpose we show that the gravitational energy and entropy are correctly obtained in the canonical frame in the two most important cases that feature horizons: Schwarzschild's black hole spacetime and de Sitter's cosmological spacetime. To compute the entropy, we use the Euclidean path integral method in the saddle point 
approximation \cite{Gibbons:1976ue}.  Since on-shell $\mathcal{R}=-8\pi GT$ we have that $\QQ=-\mathcal{D}_\alpha ( Q^\alpha - \tilde{Q}^\alpha) - 8\pi GT$, where $\mathcal{D}_\alpha$ is the metric-covariant derivative \cite{BeltranJimenez:2018vdo}. Thus we may consider the action (\ref{lq}) in the form 
\be \label{form2}
S_E = -\frac{1}{16\pi G}\int \diff^4 x \partial_\alpha\lb \sqrt{-g}( Q^\alpha - \tilde{Q}^\alpha)\rb+ \int \diff^4 x \sqrt{-g}\lp L_M-\frac{1}{2}T\rp\,.
\ee
Note that a cosmological constant does not contribute to this formula.

For the two cases of interest, very conveniently a solution with $\mathring{t}^\mu{}_\nu=0$ exists. These solutions are included in the Kerr-Schild class of metrics \cite{Debney:1969zz,Aguirregabiria:1995qz} of the form 
\be \label{ksmetric}
g_{\mu\nu} = \eta_{\mu\nu} + 2V(r)\ell_\mu\ell_\nu\,, \quad \ell_\mu\diff x^\mu = \diff t + \delta_{ij}\frac{x^i}{r}\diff x^j\,, \quad r \equiv \delta_{ij}x^ix^j\,.
\ee
Note that $\ell_\mu$ is a geodesic null vector. We can straightforwardly compute the two traces and obtain that $Q^\mu=0$ and $\tilde{Q}^\mu = -2V'(r)\ell^\mu$ (circles for the coincident gauge omitted here). In the Euclidean case we let
$\eta_{\mu\nu} \rightarrow \delta_{\mu\nu}$ and $\diff t \rightarrow -i\diff\tau$ in (\ref{ksmetric}) and can then calculate (\ref{form2}):
\ba \label{entropy}
S_E & = &  \frac{1}{16\pi G}\int \diff^4 x \partial_\alpha \tilde{Q}^\alpha =  \frac{1}{16\pi G}\int_{r=r_+} \diff^3 x \tilde{Q}^\mu n_\mu \nonumber \\
 &  =  & -\frac{1}{8\pi G}r_+^2 V'(r_+) \int_0^{\beta} \diff\tau\int \diff^2 \Omega = -\frac{\beta}{2}r_+^2V'(r_+)\,.
\ea
In the first line we used the Gauss theorem and referred to the outward unit normal as $n_\mu$, with respect to the surface of the horizon at $r=r_+$, and in the second line referred to the period of the Euclidean time $\tau$ as
$\beta$. In the case of a black hole, the temperature $1/\beta$ of the horizon at $r_+$ is given by $1/\beta=-V'(r_+)/(2\pi)$, and the above reduces to the statement of the area law $S_E=\pi r_+^2$, regardless
of the $V(r)$ (in the case of the Schwarzschild black hole, $V(r)=GM/r$). The energy 
according to the $\mathring{\tau}^\mu{}_\nu$ in the coordinates (\ref{ksmetric}) has been already investigated \cite{Aguirregabiria:1995qz} and it turns out to be equal to the mass $M$ when $V(r)=GM/r$. The reader may 
notice that in CGR we neatly avoid invoking a regulating term as well as the boundary term which is solely responsible for the entropy in the original derivation of Gibbons and Hawking \cite{Gibbons:1976ue}.  

The case of the cosmological horizon is somewhat different. The static patch of de Sitter space is described by $V=-r^2/r_+^2$, where
$r_+$ is now understood as the  cosmological horizon at which $\beta=2\pi r_+^2$. We thus obtain from (\ref{entropy}) that $S_E = -\pi r_+^2$, which again is the correct result since the energy in this case is vanishing. 
\newline
\newline

This essay was meant to substantiate the claims that in purified gravity 1) energy-momentum is described locally by a conserved, symmetric tensor, 2) the path integral formalism requires neither boundary nor regulating terms and 3) we can, in covariant terms, identify a canonical frame wherein to perhaps carry out quantisation. Though inertial frames are of paramount importance in special relativity, the idea of preferred frames is alien to the standard interpretation of GR that is formally based on the Hilbert action and conceptualised in terms of spacetime geometry. That interpretation is exceedingly beautiful but, maybe, it can to an extent mystify and obscure the more fundamental nature of the gravitational interaction. 
In some ways at least, CGR might be closer to Einstein's own view of GR than what has become conventionally established as ``Einstein's GR''.   
 \newline
 \newline
{\tiny {\bf Acknowledgments}. JBJ acknowledges support from the  {\it Atracci\'on del Talento Cient\'ifico en Salamanca} programme and the MINECO's projects FIS2014-52837-P and FIS2016-78859-P (AEI/FEDER). LH is supported by funding from the European Research Council (ERC) under the European Unions Horizon 2020 research and innovation programme grant agreement No 801781 and by the Swiss National Science Foundation grant 179740. TK would like to thank Laur J\"arv and Ott Vilson for useful comments on the manuscript. This research was funded by the Estonian Research Council grant PRG356 ``Gauge Gravity'', and is based upon
work from COST Action CA15117, supported by COST  (European Cooperation in Science and Technology).}

\end{document}